\documentclass{article}

\usepackage[utf8]{inputenc}
\usepackage[T1]{fontenc}
\usepackage{libertine}

\usepackage{xcolor}
\usepackage{graphicx}
\usepackage{svg}
\usepackage{caption}
\usepackage{parskip}

\usepackage{apacite}

\usepackage{listings}
\usepackage{xcolor}

\lstdefinelanguage{json}{
  basicstyle=\ttfamily,
  numbers=left,
  numberstyle=\tiny,
  stepnumber=1,
  numbersep=8pt,
  showstringspaces=false,
  breaklines=true,
  literate=
   *{0}{{{\color{blue}0}}}{1}
    {1}{{{\color{blue}1}}}{1}
    {2}{{{\color{blue}2}}}{1}
    {3}{{{\color{blue}3}}}{1}
    {4}{{{\color{blue}4}}}{1}
    {5}{{{\color{blue}5}}}{1}
    {6}{{{\color{blue}6}}}{1}
    {7}{{{\color{blue}7}}}{1}
    {8}{{{\color{blue}8}}}{1}
    {9}{{{\color{blue}9}}}{1}
}

\title{Refugees of the Digital Space: Platform Migration from TikTok to RedNote}
\author{Ziyue Feng, Tianjia Dong, Zheya Lei}
\date{July 2025}

\begin{document}

\maketitle

\begin{abstract}
In January 2025, the U.S. government enacted a nationwide ban on TikTok, prompting a wave of American users - self-identified as "TikTok Refugees" - to migrate to alternative platforms, particularly the Chinese social media app RedNote (Xiaohongshu). This paper examines how these digital migrants navigate cross-cultural platform environments and develop adaptive communicative strategies under algorithmic governance. Drawing on a multi-method framework, the study analyzes temporal posting patterns, influence dynamics, thematic preferences, and sentiment-weighted topic expressions across three distinct migration phases: Pre-Ban, Refugee Surge, and Stabilization.

An entropy-weighted influence score was used to classify users into high- and low-influence groups, enabling comparative analysis of content strategies. Findings reveal that while dominant topics remained relatively stable over time (e.g., self-expression, lifestyle, and creativity), high-influence users were more likely to engage in culturally resonant or commercially strategic content. Additionally, political discourse was not avoided, but selectively activated as a point of transnational engagement.

Emotionally, high-influence users tended to express more positive affect in culturally connective topics, while low-influence users showed stronger emotional intensity in personal narratives. These findings suggest that cross-cultural platform migration is shaped not only by structural affordances but also by users’ differential capacities to adapt, perform, and maintain visibility. The study contributes to literature on platform society, affective publics, and user agency in transnational digital environments.
\end{abstract}

\section{Introduction}

On January 19, 2025, the U.S. government enacted the \textit{Protecting Americans from Foreign Adversary Controlled Applications Act}, officially banning TikTok due to national security concerns. In response, a large number of American users migrated to alternative platforms, particularly the Chinese social media app \textit{Xiaohongshu}, internationally known as \textit{RedNote}. As downloads surged, users began referring to themselves as ``TikTok Refugees,'' using hashtags like \#TikTokRefugee to connect with one another and engage in dialogue with Chinese users. This sudden platform migration illustrates how digital environments serve not only as tools for connection, but as sociotechnical systems where cultural identities, geopolitical disruptions, and algorithmic governance converge.

Most of the time, social interaction emerging from connections between people and their environments is spatially situated~\cite{schatzki1991spatial}. As emerging information infrastructures, digital platforms transcend these physical constraints, enabling communicative flows across subjects and systems. Digital platforms have increasingly become critical media for global cultural transmission. The structural homology between social and networked life has given rise to what scholars term the \textit{``platform society''}, offering new perspectives on international communication.

Platforms are not only channels of information dissemination but also active selectors of which information gains influence potential. Using algorithms, social media applications construct distinct user experiences and platform identities. These algorithms classify, filter, and prioritize content to manage visibility, effectively determining who and what is surfaced on users’ recommendation interfaces~\cite{gillespie2014media}. Through everyday use, platforms generate new forms of stimulation, cultivate preferences and habits, and actively shape how individuals live, perceive, and relate, mediating our connections with the world and with one another~\cite{steinberg2019platform}.

Existing digital platform research remains predominantly Western-centric, often overlooking the regional particularities of platform services, content production, user interaction, and governance structures~\cite{steinberg2017introduction}. Studying the emergence of the ``TikTok Refugee'' phenomenon is therefore valuable, as it represents a case in which individuals voluntarily bypass regional restrictions and participate in cross-platform activities. This phenomenon offers a unique research opportunity to challenge Western-centrism, to propose a new framework for understanding how global users migrate across platforms in response to policy interventions, and to explore emergent forms of cross-cultural communication.

In addition, existing cross-cultural communication research has largely centered on top-down frameworks of content production and transmission, often overlooking bottom-up, user-driven practices of cultural negotiation and interaction within platform-mediated environments. Current research in this field primarily focuses on cultural difference, conflict, and integration, typically from a unidirectional and essentialist perspective centered on how to produce culturally translatable content. Less attention has been given to bi-directional, user-driven interactions, particularly spontaneous communicative behaviors under platform capitalism.

Drawing on Michel de Certeau’s \textit{The Practice of Everyday Life}~\cite{ahearne2010michel}, our study conceptualizes \textit{RedNote} as a ``third space'' where institutional strategies confront everyday user tactics. We examine how TikTok Refugees, operating within platform constraints, engage in daily interactions that transcend traditional cultural frameworks and enable participatory co-creation and bi-directional circulation of cultural content, offering a new framework for cross-cultural communication research.

Grounded in this theoretical perspective, the study adopts a multi-method approach-combining interaction pattern analysis, sentiment analysis, and topic modeling. By examining both group- and individual-level behaviors of TikTok Refugees on RedNote, the study systematically analyzes the dynamics and influence of platform-migration users within a newly mediated sociotechnical environment.

\section{Literature Review}

The concept of \textit{algorithmic visibility} within platformized societies has increasingly garnered scholars’ attention. From the critical perspective of platform centralism, algorithms, as structurally embedded components, embody the logic of power and control driven by capital interests. From the lens of popular culture, however, researchers are more concerned with how users construct what are known as \textit{algorithmic imaginaries} - ideas about how algorithms operate - and how these imaginaries, grounded in users’ everyday experiences, inform the production of content that aligns with and reinforces the distinctive affordances and norms of specific platforms~\cite{devito2017algorithms}. Algorithmic imaginaries can be understood as bottom-up, informal, yet relatively systematic folk theories of how algorithms function, developed by users through their everyday engagement with platforms. Chinese scholar Wang Yan and colleagues have illustrated the algorithmic imaginaries of Xiaohongshu (\textit{RedNote}) users, arguing that the platform’s algorithm fosters user experiences centered on content diversity, sincere communication, and altruistic production.

Wang Yan and colleagues also point out that even when users develop a positive algorithmic imaginary of Xiaohongshu (\textit{RedNote}), the underlying dynamics of algorithmic exploitation and user domestication remain significant and should not be ignored. This critical perspective resonates with broader tendencies toward stratification and decentralization within the digital landscape.

Building on this foundation, several critical questions emerge in the context of the recent influx of ``TikTok Refugees'': Have these digital migrants already begun to develop their own content dissemination strategies tailored to the affordances and norms of RedNote? Conversely, have platform ``natives'' started to exhibit dietary biases - habitual content preferences - toward refugee-generated content?

To address these questions, this study examines the connection patterns and interaction behaviors of TikTok Refugees on RedNote, and proposes the following research question:

\begin{quote}
\textbf{Study1:} What are the overall migration characteristics of TikTok Refugees during the platform transition, and how have subgroup structures formed on RedNote?
\end{quote}

Since Harold Lasswell introduced the concept of the \textit{``propaganda front''} in 1927~\cite{lasswell1971propaganda}, international political communication has undergone multiple transformations and expansions. It has evolved from state-led, one-directional wartime propaganda to a dual-centered interactive model involving both propagators and audiences as active agents. Changes in communication channels have played a pivotal role in this evolution.

Jürgen Habermas, in his theory of communicative action, posits that ``participants in communicative action always reach understanding within the horizon of the lifeworld. This lifeworld is composed of background assumptions that, while varied, are never radically incompatible. Such a background is the very source from which participants define their situation~\cite{habermas1985theory}.'' With the emergence of digital platforms, a globally shared communicative horizon has become a tangible reality, thereby increasing the potential and opportunities for cross-cultural exchange.

Chinese scholar Wu Fei proposed the concept of the \textit{Digital Commons}, arguing that digital spaces provide broader grounds for international communication, and that \textit{emotion} serves as a key mechanism linking users across global digital platforms~\cite{YANJ202306002}. ``Even in the presence of stark cultural differences, people can still discover commonalities through emotional practices, thereby facilitating global cooperation and consensus.'' Wu advocates incorporating emotion, values, and other subjective dimensions as central issues in international communication studies.

Empirical research in this domain has increasingly focused on emotional resonance across topics on social media platforms. For instance, Cheng Siqi and Yu Guoming, through their analysis of YouTube video comments, demonstrate that \textit{hedonic culture} - grounded in shared human emotions and values - can generate emotional resonance by evoking feelings of happiness, thereby enabling effective cross-cultural communication. In contrast, Deng Chunlin and colleagues, building on Wilson’s framework, argue that highly polarized negative emotions tend to attract more attention and visibility.

Given this, an important question arises: as a newly emerged group, what emotional tendencies do TikTok Refugees exhibit in their content production?

\begin{quote}
\textbf{Study2:} What communicative strategies and emotional expressions distinguish high-influence and low-influence users across platform phases?
\end{quote}

\section{Data Collection}

From January to June 2025, we collected user-generated content from RedNote using the keywords \textit{``tt refugee,'' ``TikTok Refugee,'' `` Tiktok ban,''} and \textit{``American Refugee,''} yielding 33{,}936 posts from 2{,}020 unique users. Among them, 1{,}299 U.S.-based users were manually identified based on location-indicative profile information, language use, and self-disclosure. Initially, data was collected between January and March 2025. However, time series analysis revealed that posting frequency had not yet stabilized during this early migration phase. To ensure a more representative and behaviorally robust dataset, we conducted an additional round of data collection in June 2025 to capture user activity from March through June.

For each identified user, we retrieved their complete homepage data, including all original posts, top-level comments (up to 50 per post), and related metadata such as likes, shares, saves, and timestamps. In total, \textbf{43,338} posts and \textbf{387,395} direct comments were collected. We did not include sub-comments (i.e., replies to comments) in the dataset, focusing only on first-level responses. This decision was based on two considerations: (1) to capture users' immediate reactions and the post's primary communicative pull, as prior research suggests first-level interactions are most indicative of attention and engagement; and (2) to manage the scope of data collection within a limited timeframe.

While comment text was primarily used for topic classification, it did not influence the construction of influence groups, which were derived exclusively from post-level engagement metrics. As such, the absence of sub-comments does not affect the reliability of our influence-based comparisons.


\section{Methodology}

\subsection{Sentiment Analysis}

To assess the emotional tone of user-generated content, we applied a pre-trained multilingual sentiment analysis model-\textit{multilingual-sentiment-analysis}~\cite{Sanh2019DistilBERTAD}-which classifies text into five sentiment levels: \textit{Very Negative}, \textit{Negative}, \textit{Neutral}, \textit{Positive}, and \textit{Very Positive}. For the purposes of statistical analysis, we converted these categorical labels into numerical values, assigning -2 to \textit{Very Negative}, -1 to \textit{Negative}, 0 to \textit{Neutral}, 1 to \textit{Positive}, and 2 to \textit{Very Positive}.

To ensure consistent emotional intensity across samples, we filtered posts with a model-generated emotional confidence score above 0.3. From each user group (high- and low-influence), we randomly sampled 400 posts ($n = 400$ per group). We then conducted chi-square tests to evaluate whether there were statistically significant differences in sentiment distribution between the groups.

\subsection{Topic Modeling}

To analyze content differences across user groups, we employed a hybrid annotation approach combining manual labeling and large language model (LLM)-based classification. Two trained annotators independently coded 20\% of the sample posts based on both the full post content and the contextual information provided by top comments (ranked by number of likes). Disagreements were resolved through discussion to establish a consistent labeling protocol and finalize the topic taxonomy.

The remaining 80\% of posts were classified using the LLM-DeepSeek model~\cite{deepseekai2024deepseekv3technicalreport}, prompted with representative examples and structured instructions derived from the manually labeled subset~\cite{sibitenda2024leveraging}. To ensure more context-aware classification, each post was paired with its top ten most-liked comments as part of the input to the model, enabling topic inference based on both original expression and audience reception.

The final taxonomy consists of 7 main categories and 15 subcategories, encompassing themes such as cultural exchange, platform adaptation, meme expression, and digital rights. To assess model reliability, a random 10\% of LLM-labeled posts were manually re-validated, yielding a Cohen’s Kappa score of 0.82, indicating substantial agreement.

\section{Study 1: Migration Phases and User Retention Patterns}
\subsubsection*{Phase Identification and Trend Analysis}


Migration behavior is typically defined as a user's reduction or complete cessation of engagement with a specific technological product in favor of another that fulfills similar needs~\cite{ye2011role}. On social media platforms, such migration is often driven by user dissatisfaction, the appeal of alternative platforms, and switching mechanisms. As a group affected by compulsory policy interventions, ``TikTok refugees" represent a unique case of proactive ``digital migration," offering valuable insights for the study of platform migration.

Building on this, the present study aims to describe the migration and retention patterns of TikTok refugees on the RedNote through a temporal analysis, providing a foundational empirical overview of cross-platform, transnational user migration.

\begin{figure}[ht]
    \centering
    \includegraphics[width=\textwidth]{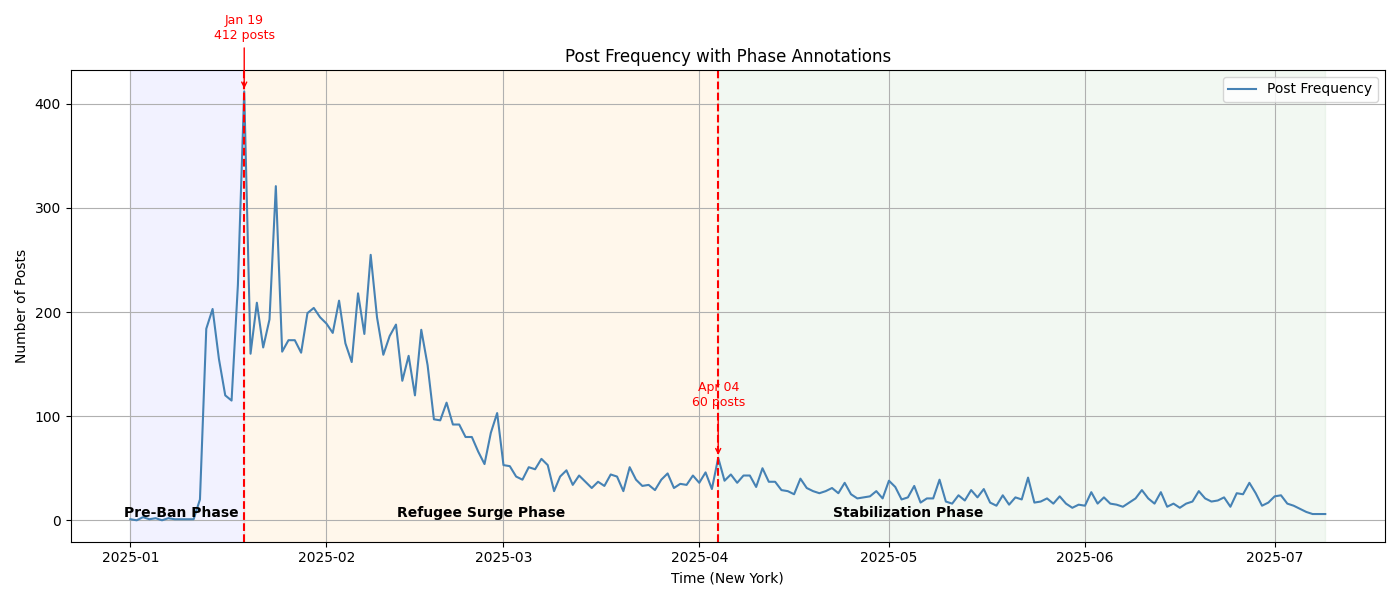}
    \caption{
    Temporal trend of post frequency on RedNote. Colored background shading enhances the visual segmentation of these phases. The timeline reflects three distinct stages of user migration and interaction, offering a macroscopic view of shifting engagement.
    }
    \label{fig:post_trend}
\end{figure}

Figure~\ref{fig:post_trend} illustrates the temporal trend of post frequency from January to June 2025, visualizing three key migration phases identified through changes in user activity: the \textit{Pre-Ban Phase}, the \textit{Refugee Surge Phase}, and the \textit{Stabilization Phase}. Post timestamps were aggregated by day in the New York timezone. The timeline begins on January 1, 2025, and phase boundaries were manually defined based on known policy events and observed shifts in activity volume.

The first critical point is January 19, 2025, the date on which the U.S. government had previously announced its plan to ban TikTok. Notably, in early January, multiple media outlets reported that RedNote (Xiaohongshu) had become the number-one downloaded app in the U.S. App Store for several consecutive days-a trend indicative of anticipatory migration behaviors already underway, which aligns with the early upward trajectory in post frequency observed in our data. On January 19 anticipated, TikTok experienced a temporary access disruption lasting approximately \textit{12} hours, further heightening public anxiety and uncertainty. Consequently, RedNote saw post volume peak at \textbf{2106} posts that day-the highest single-day count in our observation period. Following the partial restoration of TikTok services, posting frequency among ``refugee'' users on RedNote began to decline, signaling a temporary retraction and hesitation amid evolving platform dynamics.

The second key point is April 4, 2025, which records the lowest post volume (64 posts) following the peak, indicating the tail end of mass migration and the onset of a stable usage pattern. Based on these inflection points, we define the following three phases:

\begin{itemize}
    \item \textbf{Pre-Ban Phase} (before Jan 19): This period reflects early exploratory activity by a small number of users, with minimal post volume.
    
    \item \textbf{Refugee Surge Phase} (Jan 19 to Apr 4): Characterized by a steep rise and gradual decline in posting activity, this phase captures the main influx of TikTok Refugees and their adaptation process.
    
    \item \textbf{Stabilization Phase} (after Apr 4): Post frequency stabilizes at a low but consistent level, suggesting that only a subset of users remained active after the initial burst.
\end{itemize}



Across the three phases, the most frequently posted topic categories showed a high degree of consistency. Content related to \textbf{self-expression and personal storytelling}, \textbf{food, lifestyle, and fashion}, and \textbf{arts and creativity} consistently appeared among the dominant themes throughout all stages. These recurring categories suggest that users maintained a strong interest in individual expression, aesthetic production, and everyday cultural practices during the platform migration process.

In terms of posting volume, overall activity rose sharply in Phase~2 (the Refugee Surge Phase) and declined significantly in Phase~3. However, despite these fluctuations in quantity, the structure of dominant topics remained relatively stable. The recurrence of similar topics across phases indicates that users’ thematic focus did not shift substantially over time.

This relative stability in topic composition suggests that analyzing post volume alone is insufficient to understand the adaptive strategies employed by TikTok Refugees in the context of platform transition. While posting volume reflects temporal changes in user activity, it does not reveal whether users leveraged distinct content strategies to build or enhance their influence, nor how their expressive practices or interactional behaviors may have varied. Therefore, we further disaggregate users based on their influence level, in order to explore how high- and low-influence users may have differed in their content preferences, communicative tactics, and platform adaptation patterns. This allows for a more nuanced understanding of user expression and interaction dynamics in the broader context of digital migration.

\section{Study2: Communicative Strategies of High- and Low-Influence Users}
\subsection{Influence Scoring and Grouping}

From the perspective of the ``integration of production and consumption,'' users are both consumers and producers of content. Accordingly, they can be categorized along two dimensions-content production capability and content consumption demand-into four types: low capability–high demand, high capability–high demand, low capability–low demand, and high capability–low demand~\cite{GGYY202210006}. However, due to limitations in the access to the platform data, this study cannot obtain the interaction data of Tiktok refugees on RedNote, making it difficult to accurately assess their consumption demand. Therefore, the study focuses on the content production aspect and classifies Tiktok refugees into high-influence and low-influence users based on their entropy-weighted influence score  for each user based on four engagement metrics: followers, likes, comments, and saves. This method assigns weights to each metric according to its information entropy, reflecting how much each metric contributes to differentiating user influence across the population.

In particular, metrics that exhibit greater dispersion (i.e., higher information content) receive larger weights, allowing the final score to adapt to the structural characteristics of the dataset. Let $x_{ij}$ denote the normalized value of user $i$ on metric $j$, and $p_{ij}$ the corresponding proportion. The entropy of each metric is defined as:

$$H_j = -\frac{1}{\log(n)} \sum_{i=1}^{n} p_{ij} \log(p_{ij})$$

where $n$ is the number of users. The redundancy $d_j = 1 - H_j$ represents the discriminative power of metric $j$, and the weight assigned to that metric is:
$$w_j = \frac{d_j}{\sum_{j=1}^{m} d_j}$$
The final influence score for each user is calculated as a weighted sum:
$$S_i = \sum_{j=1}^{m} w_j \cdot x_{ij}$$


\subsection{Descriptive Comparison of User Groups}

\begin{figure}[ht]
  \centering
  \includegraphics[width=\textwidth]{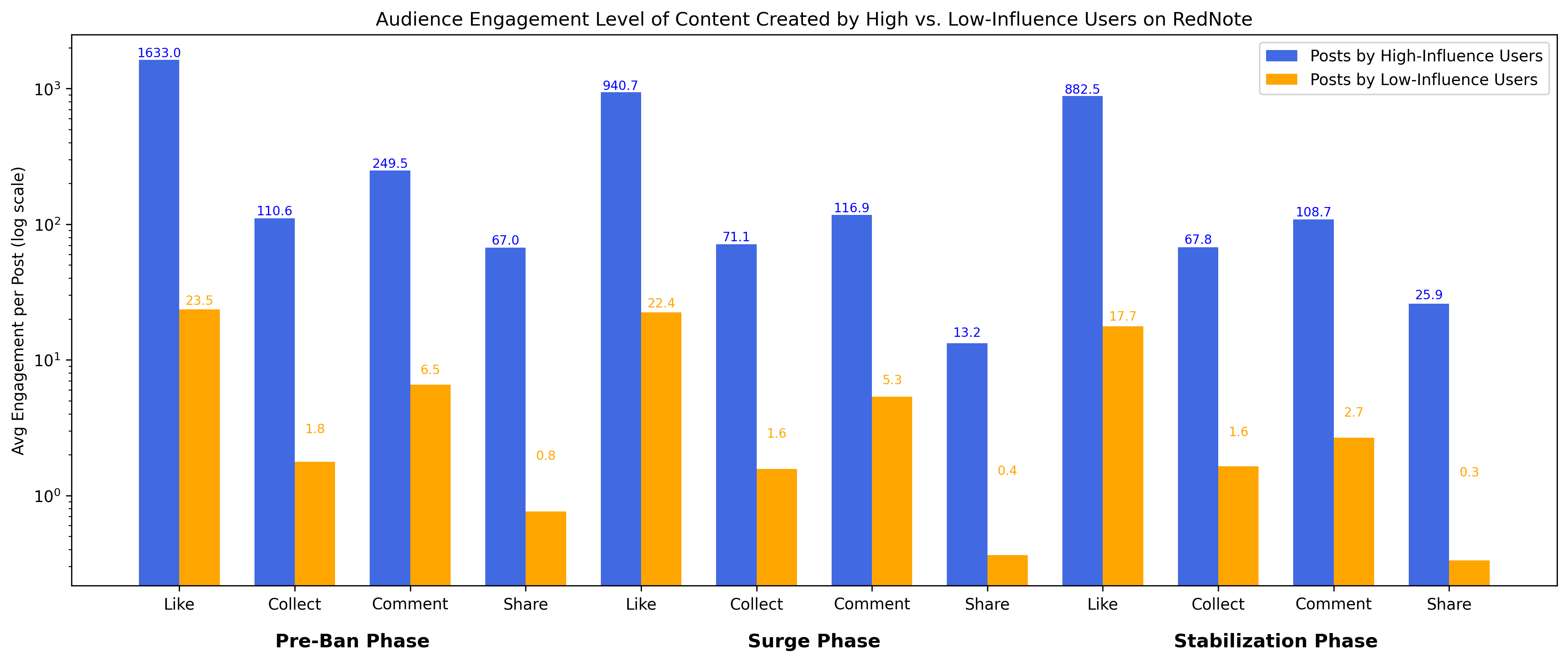}  
  \caption{Audience Engagement Level of Content Created by High vs. Low-Influence Users Per Phase}
  \label{fig:phase_engagement}
\end{figure}

Figure~\ref{fig:phase_engagement} presents a comparative overview of audience engagement levels across four interaction metrics-likes, collects, comments, and shares-for high- and low-influence users throughout the three platform migration phases. Overall, high-influence users consistently outperformed low-influence users across all dimensions, receiving exponentially more likes, comments, and shares per post. For instance, in the Pre-Ban Phase, high-influence users averaged over 1600 likes per post, compared to just 23.5 for low-influence users. This engagement gap remained evident across subsequent phases, underscoring the disproportionate visibility and reach of high-influence users.

While engagement levels for high-influence users declined slightly from the Pre-Ban to Stabilization Phase-likely reflecting an overall decrease in platform activity-the relative difference between the two user groups persisted. In contrast, low-influence users maintained consistently low levels of engagement across all three phases, indicating limited audience expansion or algorithmic traction. These trends suggest a structurally unequal interaction landscape, where influence correlates strongly with sustained engagement advantages over time.

\begin{figure}[htbp]
    \centering
    \includegraphics[width=1.0\linewidth]{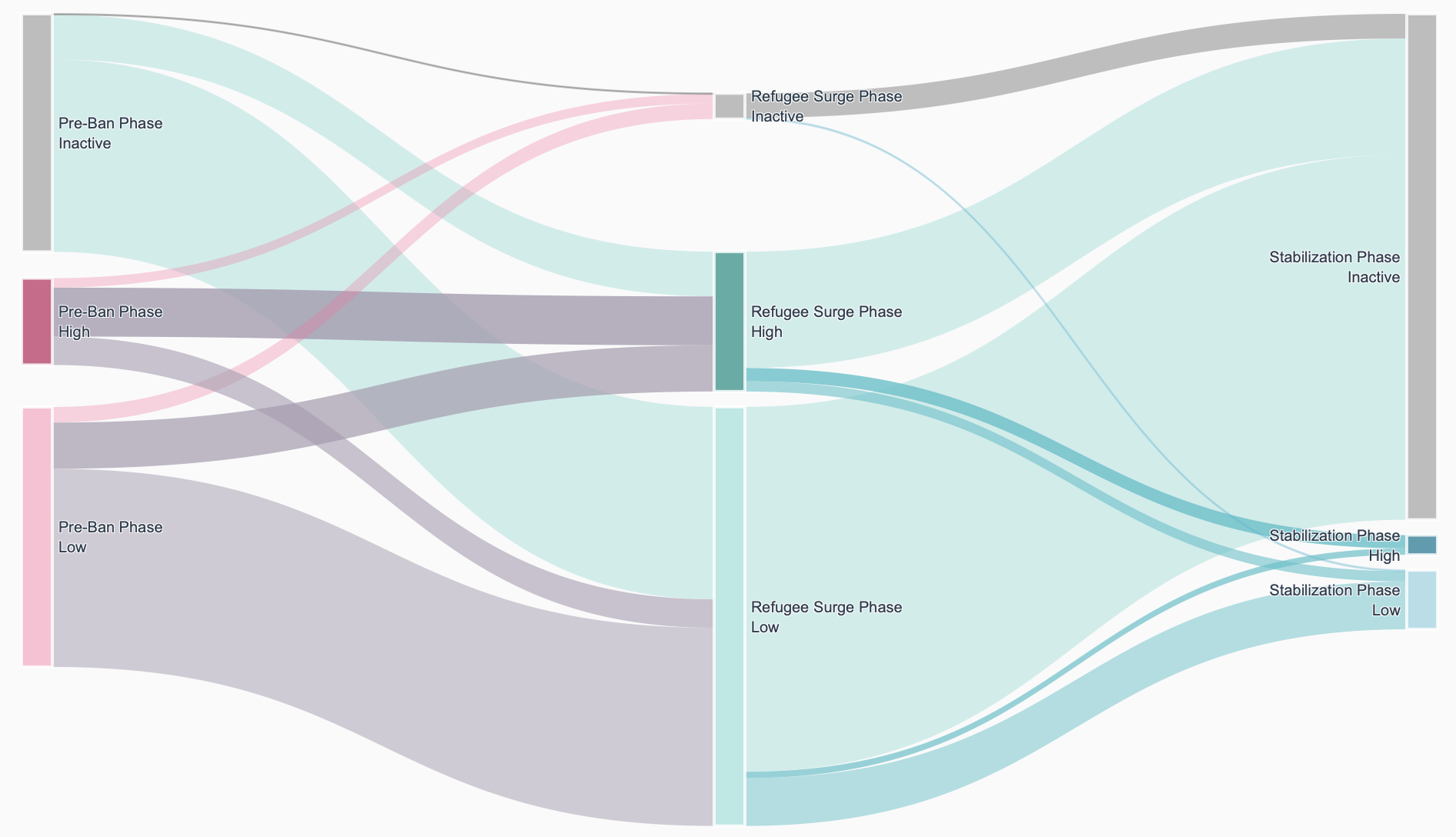}
    \caption{
    User trajectory across platform migration phases: Each phase-Pre-Ban, Refugee Surge, and Stabilization-is divided into three user types based on activity level: Inactive(users with fewer than 3 posts in a phase), Low influence(bottom 75\% by influence score), and High influence(top 25\%). The Sankey diagram visualizes user transitions between these categories across phases, revealing patterns of drop-off, influence emergence, and stabilization.
    }
    \label{fig:Sankey_trend}
\end{figure}

As shown in Figure~\ref{fig:Sankey_trend}, a substantial portion of users transitioned into the \textit{inactive} category by the Stabilization phase. This observation aligns with the overall downward trend in user activity presented in Figure~\ref{fig:post_trend}. Such a pattern highlights the inherent challenges of cross-platform migration. The majority of users failed to sustain their engagement over time, suggesting structural constraints in adapting to new digital environments. Prior research indicates that user disengagement is a common phenomenon in platform shifts, where frictional factors-such as unfamiliar interface designs, algorithmic opacity, and differences in platform culture-impede the continuity of prior interaction habits~\cite{ittefaq2025digital}.

At the same time, we observe a notable path continuity among high-influence users: many who were highly influential during the Pre-Ban phase remained so in the Refugee Surge phase. This suggests that a subset of users successfully transferred their influence across platforms. More interestingly, a portion of low-influence users in the first phase transitioned into the high-influence group in the second, indicating that influence is not strictly path-dependent. Instead, certain users appear to have strategically cultivated their visibility and engagement on RedNote.

These upward transitions imply the presence of adaptive strategies tailored to the affordances of the new platform. As noted by Ekinci et al.~\cite{ekinci2025dark}, content creators often respond to platform-specific dynamics by modifying their communicative styles, aesthetic choices, and engagement tactics. This capacity for adaptation reflects a form of user agency within algorithmically mediated environments. The flow patterns observed in Figure~\ref{fig:post_trend} not only confirm the presence of intentional communication strategies among TikTok Refugees but also illustrate how users learn, adapt, and innovate in order to rebuild their influence networks amidst sociotechnical displacement.



\subsection{Topic Preferences and Strategic Content Adaptation}

\begin{figure}[ht]
  \centering
  \includegraphics[width=\textwidth]{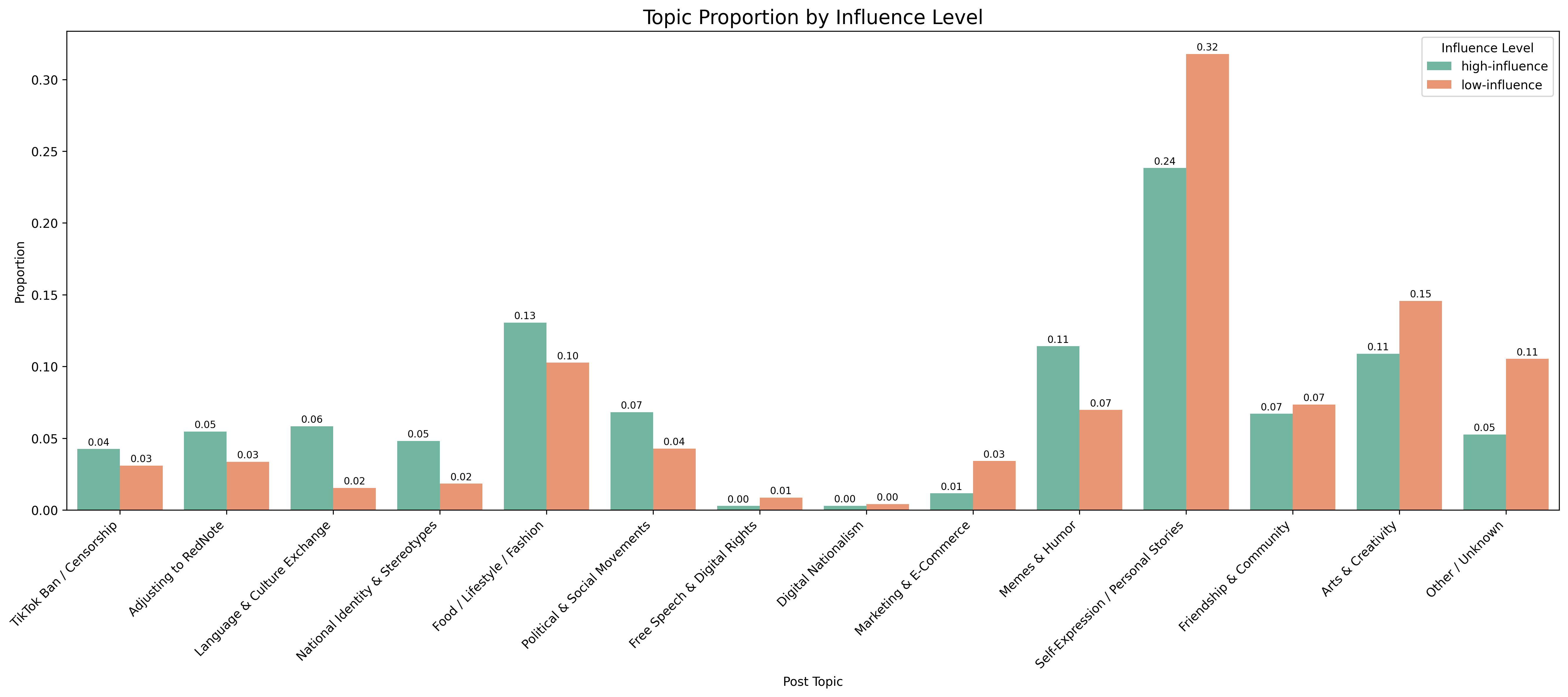}  
  \caption{Topic Proportion by Influence Level}
  \label{fig:phase-engagement}
\end{figure}

We compare the topic distribution of posts by high- and low-influence users (see Figure~\ref{fig:phase-engagement}). While both groups show a shared preference for a small set of dominant topics, such as \textit{Self-Expression, Personal Stories}, \textit{Friendship \& Community}, and \textit{Arts \& Creativity}, several differences in topic emphasis are noteworthy.

Low-influence users show a substantially higher proportion of posts related to \textit{Self-Expression, Personal Stories}, accounting for over 30\% of their content, compared to 24\% for high-influence users. They also demonstrate a stronger focus on \textit{Arts \& Creativity} and \textit{Other, Unknown} categories. In contrast, high-influence users contribute relatively more content on topics such as \textit{Food \& Lifestyle \& Fashion}, \textit{Political \& Social Movements}, and \textit{Language \& Culture Exchange}. These differences suggest that high-influence users are more likely to engage in topical content that signals cultural awareness, lifestyle aesthetics, or socio-political relevance-topics that may be more conducive to attracting visibility, resonance, or algorithmic promotion. Low-influence users, on the other hand, tend to concentrate more on personal narratives and informal creative expression, which may reflect a more individualized or less strategic approach to content production.

While the above comparison reveals overall topic preferences between high- and low-influence users, it does not capture how these preferences evolve across different migration phases. In particular, it remains unclear what topics were most prominent or strategically leveraged during each phase. To address this, we further examine the top categories among high-influence users based on engagement entropy. This allows us to identify the most attention-attracting topics within each phase and to explore how influential users potentially adapt their content strategies over time.

\begin{figure}[ht]
  \centering
  \includegraphics[width=\textwidth]{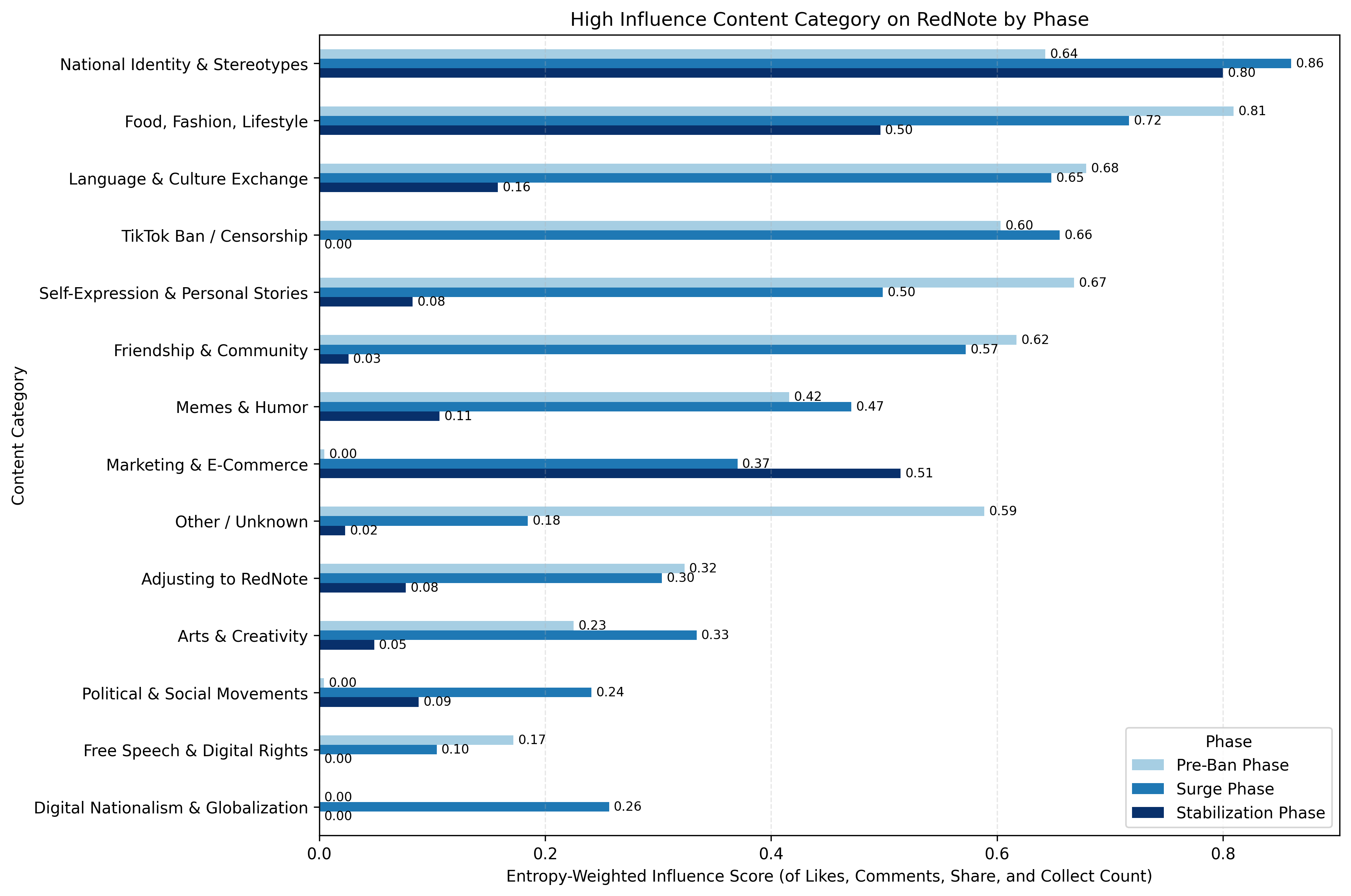}  
  \caption{Top categories among high-influence users based on engagement entropy.}
  \label{fig:high_influence_phase}
\end{figure}

Within Figure~\ref{fig:high_influence_phase}, Among high-influence users, several content categories exhibit distinct temporal trajectories in their entropy-weighted influence scores. \textit{Food / Fashion / Lifestyle} consistently ranks among the highest-performing categories across all three phases, with scores increasing from 0.50 in the Pre-Ban Phase to 0.81 in the Stabilization Phase. This upward trend reflects the strategic use of lifestyle-oriented content to sustain visibility and engagement throughout the migration process. 

\textit{Language \& Culture Exchange} demonstrates a similar increase, reaching 0.68 in the Stabilization Phase. This pattern indicates a growing emphasis on intercultural communication and identity negotiation, reinforcing the role of cross-cultural discourse in influence consolidation. In contrast, \textit{Self-Expression / Personal Stories} peaks during the Refugee Surge Phase (0.67) and declines in the Stabilization Phase (0.50), indicating a shift from personal narration to more outward-facing content as user networks mature and platform expectations evolve. \textit{Marketing \& E-Commerce} shows a marked rise from 0.00 in the Pre-Ban Phase to 0.51 in the Stabilization Phase. This shift reflects the increasing integration of commercial strategies into content production, corresponding to the platform’s growing commercialization and the professionalization of influential users.


\subsection{Affective Dynamics of Thematic Content}

In the context of social media, emotion functions not only as a core component of user expression, but also as a critical mechanism linking content visibility, algorithmic decision-making, and cross-cultural identity construction. Prior research has shown that within platformized communication infrastructures, affect serves as a mediating variable in the ``algorithmic value chain,'' significantly shaping patterns of content amplification and community formation~\cite{papacharissi2015affective}. Especially in contexts of cross-cultural platform migration, the emotional intensity and polarity embedded in different content categories can offer valuable insight into how users navigate personal expression, cultural adaptation, and public participation.

Building on this foundation, we introduce the metric of \textit{sentiment-weighted topic intensity} to systematically assess the emotional profiles of different content categories. This metric incorporates both the sentiment polarity of each post (predicted by a pretrained sentiment classification model, ranging from $-2$ to $+2$). In this analysis, we focus on unweighted average sentiment to establish a baseline comparison of emotional tone across topics.

\begin{figure}[ht]
  \centering
  \includegraphics[width=\textwidth]{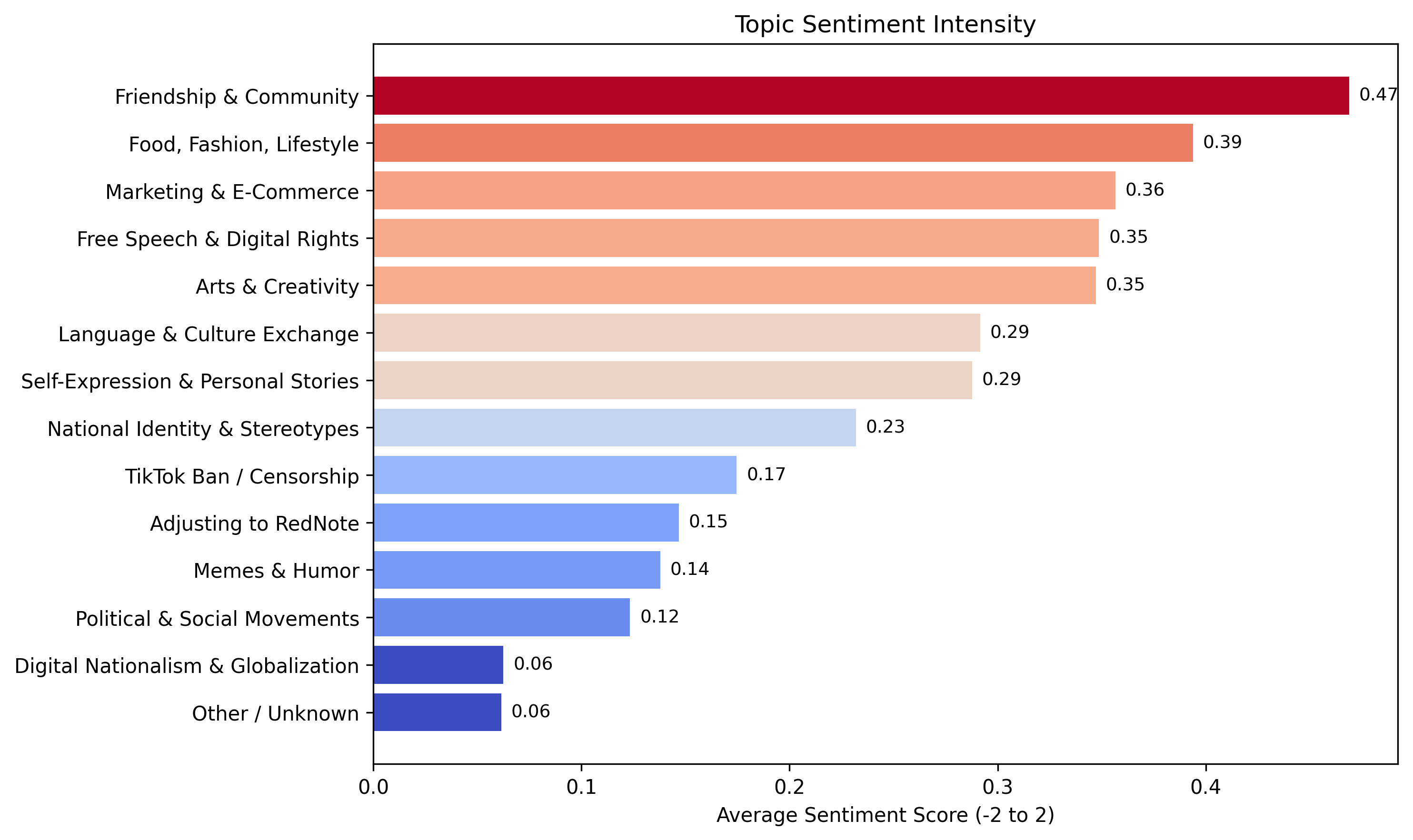}  
  \caption{Top categories among high-influence users based on engagement entropy.}
  \label{fig:fig6}
\end{figure}

As shown in Figure~\ref{fig:fig6}, content related to \textit{Friendship \& Community} exhibits the highest average sentiment score ($0.47$), followed by \textit{Food / Fashion / Lifestyle} ($0.39$) and \textit{Marketing \& E-Commerce} ($0.36$). These topics are characterized by strong positive emotional valence and are often associated with themes of connection, aspiration, and enthusiasm. Other categories such as \textit{Free Speech \& Digital Rights} and \textit{Arts \& Creativity} also display elevated sentiment, reflecting expressive autonomy and creative affirmation.

Conversely, more politically or structurally oriented topics-including \textit{Political \& Social Movements}, \textit{Digital Nationalism \& Globalization}, and \textit{TikTok Ban / Censorship}-demonstrate lower average sentiment scores. These patterns suggest that emotionally positive topics are more affectively charged and community-driven, while emotionally neutral or negative topics tend to engage with more contentious, ideological, or critical discourse.

The distribution of sentiment intensity across topics underscores the affective stratification of cross-cultural communication on platformed media. Emotionally rich topics contribute not only to user engagement but also to the formation of affective publics, indicating that emotional tone plays a key role in shaping both user behavior and platform dynamics.

\begin{figure}[ht]
  \centering
  \includegraphics[width=\textwidth]{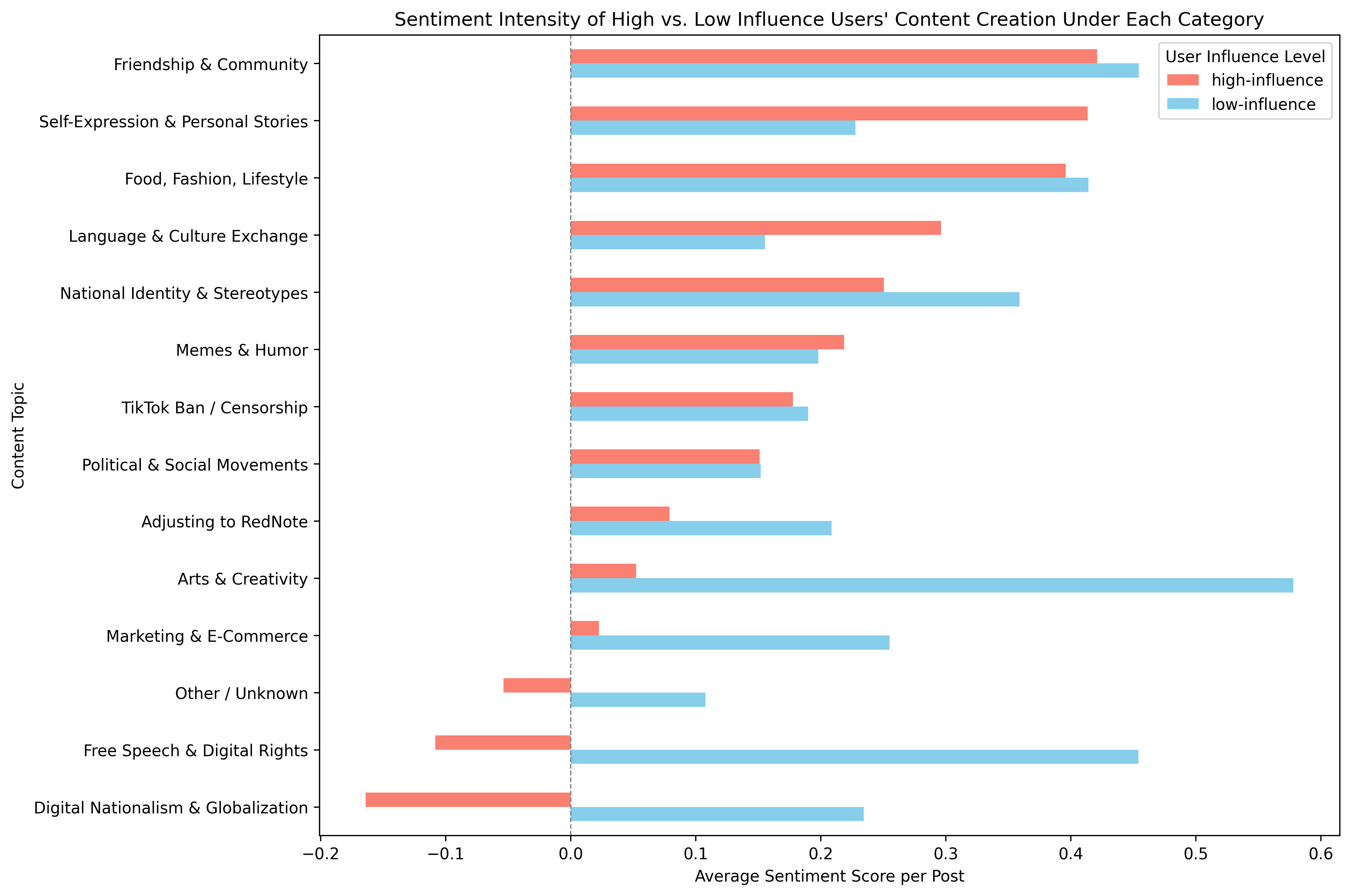}  
  \caption{Sentiment Intensity of High vs. Low Influence Users' Content Creation Under Each Category}
  \label{fig:fig7}
\end{figure}


To further explore the relationship between user influence level and emotional expression, we analyze the average sentiment scores for representative topics across high- and low-influence users. Figure~\ref{fig:fig7} presents the sentiment distribution across categories, highlighting how emotional tone varies between groups. This enables us to trace possible affective trajectories and communicative strategies. Notably, low-influence users consistently demonstrate higher sentiment scores across most content topics, particularly in emotionally expressive categories such as \textit{Self-Expression \& Personal Stories}, \textit{Friendship \& Community}, and \textit{Arts \& Creativity}. This pattern reinforces previous findings that low-influence users rely more heavily on affectively charged, personal, or community-centered content-often with more enthusiastic or optimistic tone. For instance, their average sentiment score for \textit{Arts \& Creativity} exceeds 0.55, compared to just 0.07 for high-influence users.

In contrast, high-influence users show comparatively neutral or even negative emotional tendencies in certain topics. For example, their posts in \textit{Digital Nationalism \& Globalization} and \textit{Free Speech \& Digital Rights} exhibit negative average sentiment scores (around -0.18 and -0.1 respectively), indicating more critical or confrontational tones. These topics are more ideologically charged and thus attract more analytical or oppositional framing, particularly among users aiming to provoke or initiate public discourse.

In political or socio-technical domains such as \textit{TikTok Ban / Censorship} and \textit{Political \& Social Movements}, both user groups show relatively low sentiment scores, but high-influence users tend to express more reserved emotional tones, perhaps indicating calculated neutrality or strategic ambiguity. Meanwhile, low-influence users present moderately higher scores, suggesting a more emotionally invested response to external policy shocks.

This comparison reveals a notable divergence in emotional communication strategies. High-influence users appear more likely to deploy emotionally neutral or negative tone in politicized or ideological contexts, while adopting selectively positive tone in lifestyle-oriented or commercially viable topics. Low-influence users, by contrast, sustain higher affective intensity across nearly all topics, signaling a more expressive but less targeted engagement pattern.




\section{Discussion}

From the perspective of the ``integration of production and consumption,'' platform-level observations suggest that platform migration behaviors driven by political and policy factors tend to be unstable. Established usage habits exert a strong pull, often prompting users to return to their original platforms. From the user perspective, consistent with prior research, interaction remains a key factor influencing user retention on platforms.

An analysis of the overall content themes published by Tiktok refugees indicates that they are not unfamiliar with RedNote. Upon entering the new platform, users already exhibit some understanding of its preferences. Their posts related to entertainment, fashion, and personal life align with the general characteristics of RedNote. This, to some extent, demonstrates that platform societies are increasingly embedded in our daily lives.

Admittedly, this study has its limitations. It primarily examines Tiktok refugees’ platform behaviors from a practical perspective, without sufficient attention to their psychological dynamics. Existing qualitative studies on RedNote users have revealed a paradoxical mindset in which algorithmic resistance and cooperation coexist-a phenomenon researchers describe as ``complicit silence'' and ``silent complicity.'' Future research may combine qualitative methods to explore how Tiktok refugees who remain on RedNote negotiate with the platform and create content and consumption environments that better reflect their personal preferences.

Furthermore, an analysis of content production by high- and low-influence users reveals that political topics are not necessarily ``taboo'' in the realm of international communication, but may serve as important points of interaction between users. This suggests that international communication should distinguish between two parallel paths: official media dissemination and personal user communication. In the latter, it may be beneficial to experiment with more diverse and bold narrative strategies.

In addition, findings related to users’ cultural closure index indicate that-even when Tiktok refugees engage on foreign social media platforms-their content production and interaction tend to reflect a preference for their native (or regional) culture. This stands in contrast to more optimistic views found in previous research on Tiktok refugees. Therefore, it is important to avoid overly idealistic assumptions about the extent to which these users assimilate or engage across cultural boundaries.

\bibliographystyle{apacite}
\bibliography{ref.bib}

\appendix

\section{Users, Posts, Comments Data Structure (JSON Sample)}

\begin{lstlisting}[language=json, caption={Sample User Entry}]
{
  user_id: 6785219f000000xxxxxxxxxx,
  nickname: anonymous user,
  gender: xxx,
  avatar: xxx,
  desc: xxx,
  ip_location: xxx,
  follows: 141,
  fans: 2299,
  interaction: 17892,
  tag_list: {location: UnitedStates},
  last_modify_ts: 1740904878931,
  post_count: 59.0,
  like_count: 17349.0,
  share_count: 332.0,
  collect_count: 527.0,
  comment_count: 5237.0,
  avg_likes: 294.05,
  avg_collects: 8.93,
  avg_comments: 88.76,
  avg_shares: 5.63,
  influence_score: 0.00428,
  influence_level: high-influence
}
\end{lstlisting}

\begin{lstlisting}[language=json, caption={Sample Post Entry}]
{
  note_id: 687226d8000000xxxxxxxxxx,
  type: normal,
  title: xxx,
  desc: xxx,
  video_url: xxx,
  time: 1752311512000,
  last_update_time: 1752312082000,
  user_id: 67856058000000xxxxxxxxxx,
  nickname: anonymous user,
  liked_count: 61,
  collected_count: 3,
  comment_count: 8,
  ip_location: xxx,
  sentiment: Very Negative,
  score: 0.4024,
  category: 5b
}
\end{lstlisting}

\begin{lstlisting}[language=json, caption={Sample Comment Entry}]
{
  comment_id: 68711d75000000xxxxxxxxxx,
  create_time: 1752243574000,
  ip_location: xxx,
  content: xxx,
  user_id: anonymous user,
  sub_comment_count: 1,
  last_modify_ts: 1752289615552,
  like_count: 1
}
\end{lstlisting}

\section{System and User Prompt for Post Categorization}

\begin{lstlisting}[language=Python, caption={System Prompt and User Prompt Script}]
SYSTEM_PROMPT = '''
You are a cross-cultural communication researcher analyzing social media posts on RedNote related to the recent migration of foreign TikTok users to the RedNote platform. Your task is to categorize each post into the most relevant category based on its content and comments.

### Categories:
1. Platform Migration & Adaptation
  1a. About TikTok Ban / Censorship
  1b. About Adjusting to RedNote
2. Cross-Cultural Communication & Identity
  2a. Language & Cultural Exchange
  2b. National Identity & Stereotypes
  2c. Food, Lifestyle & Fashion
3. Social & Political Discourse
  3a. Political & Social Movements
  3b. Free Speech & Digital Rights
  3c. Digital Nationalism & Globalization
4. Marketing & E-Commerce
5. Creative Expression & Social Interaction
  5a. Memes & Humor
  5b. Self-Expression & Personal Stories
  5c. Friendship & Community Building
6. Arts & Creativity
7. Unknown / Not Relevant
'''

def generate_user_prompt(post):
    note_id = post.get('note_id')
    comments = comments_by_note.get(note_id, [])[:10]
    comment_texts = [f'{i+1}. {c.get('content', '').strip()}'
                     for i, c in enumerate(comments) if c.get('content', '').strip()]
    comment_section = '\n'.join(comment_texts) if comment_texts else 'No comments available.'
    return f'''
### Post to Categorize:
Title: {post.get('title', 'No Title')}
Description: {post.get('desc', 'No Description')}

### Top 10 Comments:
{comment_section}

Please return only the category label, e.g., 1a or 3b. Do not include any other text.
'''
\end{lstlisting}

\end{document}